\documentclass[english,prl,twocolumn, aps,amssymb,footinbib,showpacs]{revtex4-1}
\usepackage[T1]{fontenc}
\usepackage{amsmath}
\usepackage{amssymb}
\usepackage{graphicx}
\usepackage{braket}
\usepackage{babel}
\usepackage{soul}
\usepackage[normalem]{ulem} 

\makeatletter

\@ifundefined{textcolor}{}
{
 \definecolor{BLACK}{gray}{0}
 \definecolor{WHITE}{gray}{1}
 \definecolor{RED}{rgb}{1,0,0}
 \definecolor{GREEN}{rgb}{0,1,0}
 \definecolor{BLUE}{rgb}{0,0,1}
 \definecolor{CYAN}{cmyk}{1,0,0,0}
 \definecolor{MAGENTA}{cmyk}{0,1,0,0}
 \definecolor{YELLOW}{cmyk}{0,0,1,0}
 }

\usepackage{bm}\@ifundefined{definecolor}{\usepackage{color}}{}

\usepackage[T1]{fontenc}
\usepackage[latin9]{inputenc}
\usepackage{textcomp}
\usepackage{color}

\usepackage{babel}

\makeatother

\begin{document}




\title{Using a Recurrent Neural Network to Reconstruct Quantum Dynamics \\ of a Superconducting Qubit from Physical Observations}

\author{E. Flurin$^{1,2,4}$}
\email{emmanuel.flurin@cea.fr}
\author{L. S. Martin$^{1,2}$}
\author{S. Hacohen-Gourgy$^{1,2,3}$}
\author{I. Siddiqi$^{1,2}$}
\affiliation{$^{1}$Department of Physics, University of California, Berkeley, CA 94720, U.S.A.\\$^2$Center for Quantum Coherent Science, University of California, Berkeley CA 94720, USA.\\$^3$ Department of Physics, Technion - Israel Institute of Technology, Haifa 32000 Israel.\\$^4$Quantronics Group, SPEC, CEA, CNRS, Universit\'e Paris-Saclay, CEA-Saclay, 91191 Gif-sur-Yvette, France.}

\begin{abstract}


At its core, Quantum Mechanics is a theory developed to describe fundamental observations in the spectroscopy of solids and gases. Despite these practical roots, however, quantum theory is infamous for being highly counterintuitive, largely due to its intrinsically probabilistic  nature. Neural networks have recently emerged as a powerful tool that can extract non-trivial correlations in vast datasets. They routinely outperform state-of-the-art techniques in language translation, medical diagnosis and image recognition. It remains to be seen if neural networks can be trained to predict stochastic quantum evolution without \textit{a priori} specifying the rules of quantum theory.
Here, we demonstrate that a recurrent neural network can be trained in real time to infer the individual quantum trajectories associated with the evolution of a superconducting qubit under unitary evolution, decoherence and continuous measurement from raw observations only. The network  extracts the system Hamiltonian, measurement operators and physical parameters. It is also able to perform tomography of an unknown initial state without any prior calibration. This method has potential to greatly simplify and enhance tasks in quantum systems such as noise characterization, parameter estimation, feedback and optimization of quantum control.

\end{abstract}
\date{\today}
\maketitle

\begin{figure}[!ht]
\includegraphics[width=0.5\textwidth]{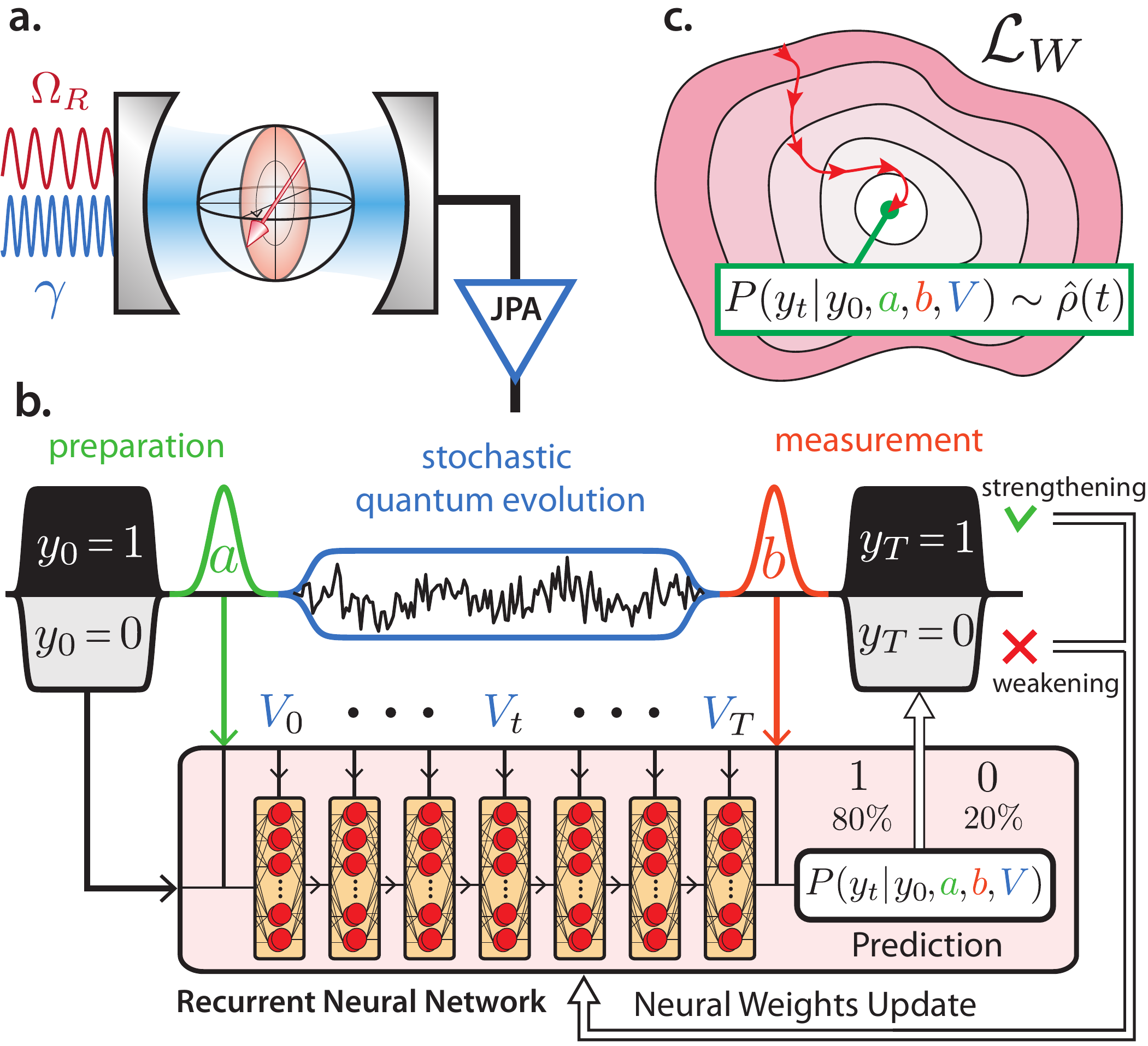}
\caption{\textbf{Recurrent Neural Network training from raw data set} \textbf{a.} Schematic of the superconducting qubit dispersively coupled to a microwave cavity monitored by a high quantum efficiency Josephson Parametric Amplifier (JPA). The qubit is simultaneously driven on resonance at a Rabi rate $\Omega_R$ and dispersively monitored with a strength $\gamma$ near the cavity resonance frequency. \textbf{b.} Data collected from the experimental system, comprising preparation, measurement outcomes and continuous measurement record of the qubit, are directly streamed to a RNN, which provides a prediction of the measurement outcome. The weights of the RNN are updated at each iteration through a stochastic gradient descent. \textbf{c.} The stochastic gradient descent aims at minimizing the cross-entropy loss function $\mathcal{L}_W$ which evaluates the distance between the prediction and the measurement outcome.}
\end{figure}

Quantum mechanics breaks dramatically with classical intuition, contradicting determinism and introducing many highly counterintuitive concepts, such as contextuality, non-classical correlations and the uncertainty principle. Despite its abstract mathematical framework, quantum mechanics can be formulated operationally as an extended information theory  \cite{Chiribella2011}, where the physical system is treated as a black box in which preparation and measurement combine to give the probabilities of experimental outcomes. The physical parameters are then estimated by averaging measurement outcomes on a large ensemble.

The time evolution of the state of an isolated quantum mechanical system is governed by the Schr\"odinger equation. However realistic system cannot be isolated perfectly, and the coupling to an environment brings about qualitatively different behavior that cannot be accounted for via the Schr\"odinger equation alone. If the system is monitored continuously, the dynamics of the system is perturbed by the inevitable back-action induced by measurement. Although the system's evolution under measurement is stochastic,
 the measurement record faithfully reports the perturbation of the system with respect to the unperturbed coherent evolution. Consequently, the observer's knowledge of the wave-function can be updated using quantum filtering - the extraction of quantum information from a noisy signal. The stochastic time evolution of the wave function is the so called quantum trajectory.
Under certain approximations, this task can be performed by integrating the stochastic quantum master-equation, provided that the Hamiltonian, dissipation and measurement operators are precisely calibrated \cite{Murch2013,Weber2014,Hacohen2016,Ficheux2018}.

On the other hand, Recurrent Neural Networks (RNN) are a powerful class of machine learning tools able to extract hidden correlations from large datasets \cite{Schuster1997}. They are most commonly applied to time-binned data, and as such achieve excellent performance on difficult problems such as language translation \cite{Mikolov2010} and speech recognition \cite{Graves2013}. RNN training is driven by examples and performed without specifying dictionaries or linguistic rules. Interestingly, quantum filtering \cite{Bouten2007} can be seen as a similar task in which noisy experimental signals must be translated  into meaningful quantum information. Last year, various architectures of neural networks have been used in the realm of quantum physics for the prediction the theoretical quantum behavior of strongly correlated phases of matter \cite{Torlai2018,Wang2016,Carrasquilla2017,van2017,Carleo2017}, the design of efficient quantum error correction code  \cite{Fosel2018}, the decoding of large topological error correcting codes \cite{Torlai2017,Krastanov2017,Baireuther2018} and the optimization of dynamical decoupling schemes for quantum memories \cite{August2017}.

In this Letter, we show that neural networks can be trained to predict stochastic quantum evolution from raw observation without specifying quantum mechanics a priori. We demonstrate that the RNN reproduces the stochastic quantum evolution for a continuously monitored superconducting qubit under a Rabi Hamiltonian. Rather than providing a black-box model, we use the neural network to 
robustly extract all physical parameters required for quantum filtering. 
 Moreover, while RNNs are temporally oriented, they are routinely trained both in the forward and backward time ordering, so that the network may exploit both past and future information. In the present application, the use of past and future continuous measurement outcomes improves the estimation accuracy of quantum trajectories at a given time through a process called quantum smoothing  \cite{Guevara2015,Tsang2009}. We train a bidirectional RNN to perform forward-backward analysis of trajectories, enabling quantum smoothing of predictions and the faithful tomography of an unknown initial state. 
By treating preparation and measurement on the same footing, the RNN structure highlights the time symmetry underlying the stochastic quantum evolution.

\section{Experimental system}
Our experiment consists of a superconducting transmon qubit \cite{Koch2007} dispersively coupled to a  superconducting waveguide cavity \cite{Paik2011}. In the interaction picture and rotating wave approximation, our system is described by the Hamiltonian $H=H_\mathrm{int}+H_\mathrm{R}$,
\begin{equation}
H_\mathrm{int}=\dfrac{ \hbar \chi}{2} a^\dagger a\ \sigma_Z,\\ 
\end{equation}
\begin{equation}
H_\mathrm{R}=\frac{\hbar \Omega_R}{2} \sigma_X
\end{equation}
where $\hbar$ is the reduced Plank's constant, $a^\dagger (a)$ is the creation (annihilation) operator for the cavity mode, and $\sigma_{X,Y}$ are qubit Pauli operators. $H_R$ describes a microwave drive at the qubit transition frequency which induces unitary evolution of the qubit state characterized by the Rabi frequency $\Omega_R$. $H_\mathrm{int}$ is the interaction term, characterized by the dispersive coupling rate $\chi=-2\pi \times 0.18\ \mathrm{MHz}$. This term describes a qubit state-dependent frequency shift of the cavity, which we use to perform quantum state measurement of our qubit. The cavity is coupled to the transmission line at a rate  $\kappa=2\pi \times 7.2\ \mathrm{MHz}$. A microwave tone that probes the cavity near its resonance frequency will acquire a qubit state-dependent phase shift.  If the measurement tone is very weak, quantum fluctuations of the electromagnetic mode fundamentally obscure this phase shift, resulting in a partial or weak measurement of the qubit state \cite{Murch2013}. We use a near-quantum-limited parametric amplifier \cite{Hatridge2011} to amplify the quadrature of the reflected signal which is proportional to the qubit state-dependent phase shift. After further amplification, we digitize the signal in $40\ \mathrm{ns}$ time steps, yielding a measurement record $V_t$. 

We begin each run of the experiment by heralding the ground state of the qubit using the above readout technique. We then prepare the qubit along one of the 6 cardinal points of the Bloch sphere by applying a preparation pulse. Next, a  measurement tone at the cavity frequency of $6.666\ \mathrm{GHz}$ continuously probes the cavity for a variable time $T$ between $0$ and $4\ \mathrm{\mu s}$ , which weakly measures the qubit in the $\sigma_Z$ basis. Concurrently, we apply the Rabi Hamiltonian $H_R$. Finally, we apply pulses to perform qubit rotations and a projective measurement, yielding a single shot measurement of a desired qubit operator $\sigma_X$, $\sigma_Y$ or $\sigma_Z$.

\begin{figure*}
\includegraphics[width=1\textwidth]{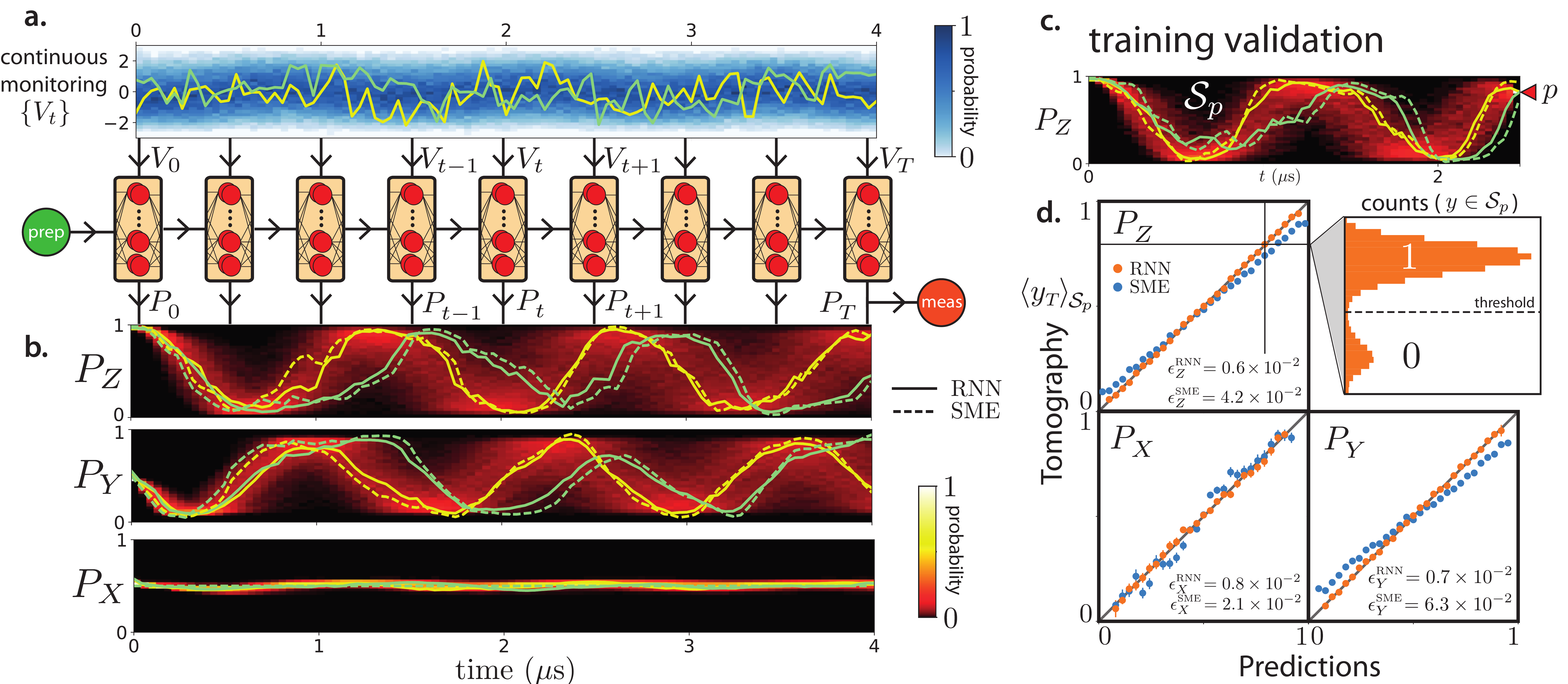}
\caption{\textbf{RNN prediction of the quantum evolution} \textbf{a}. Blue-scale histogram of the normalized measurement records extracted from the experiment, traces plotted in color show representative instances.  \textbf{a}. Red-scale histograms of RNN prediction for the measurement basis b=$X$,$Y$ and $Z$ in the driven case, beginning from $y_0=1$ in the preparation basis a=$X$. Traces plotted in color show representative instances. \textbf{c}. Training validation;  Ensemble of RNN prediction $\mathcal{S}_p$ leading to $p=0.85$ at $T=2.5\ \mathrm{\mu s}$ indicated by the red maker.  \textbf{d}. Comparison of the RNN prediction with the tomography ---averaged measurement outcome $y_t$,  \textbf{Inset} -  Ensemble of projective measurement for the predicted ensemble $\mathcal{S}_p$.}
\label{fig:trajectories}
\end{figure*}

\section{Quantum trajectories}

To allow the neural network to operate as generally as possible, we formulate system inputs and outputs symmetrically, and avoid passing it objects such as a wave function that encode information about the structure of quantum theory. The role of the wave-function in quantum mechanics is to provide the probability of a measurement outcome $y_t$ given the preparation and evolution of the system at earlier times $P(y_t|y_0)$. In the case of a continuously monitored quantum bit, the preparation and measurement outcome are each a binary variable $y_0,y_t\in \{0,1\}$  extracted through a projective readout performed at the initial and final times respectively; the preparation and measurement configurations, labeled $a$ and $b$, encode microwave pulses performing qubit rotations for state preparation and tomography respectively in the $X$, $Y$ and $Z$ basis. The stochastic measurement record $\{V_t\}$ is collected with a high quantum efficiency parametric amplifier during the qubit evolution. 
Quantum trajectory theory describes how an observer's state of knowledge evolves given a measurement record \cite{Gambetta2008}. Therefore, quantum trajectories are specified by $P(y_t|y_0,a,b,V_0...V_t)$, the probability of measuring the outcome $y_t$ with the measurement parameter $b$ given the initial measurement $y_0$ in the preparation parameter $a$ and the stochastic measurement outcome up to a time $t$. Tracking this quantum evolution can be understood as a translation of the measurement records into a quantum state evolution. Fig.2 a. shows the distribution of measurement records obtained for the preparation setting ($y_0=0$, a=$Z$). 

Quantum trajectories are typically extracted from continuous measurement by integrating the stochastic master equation (SME) governing the evolution of the density matrix $\rho_t$
\begin{equation}
d \rho_t=\underbrace{(i[H_R,\rho_t]+\mathcal{L}[\sqrt{\frac{\gamma}{2}}\sigma_Z]\rho_t) }_\text{dissipative evolution}dt+\underbrace{\sqrt{\eta}\mathcal{H}[\sqrt{\frac{\gamma}{2}}\sigma_Z]\rho_t
}_\text{backaction}dw_t.
\end{equation}
where $\mathcal{L}$ is the Lindblad superoperator describing the qubit dephasing induced by the measurement of strength $\gamma$, $\mathcal{H}$ is a measurement superoperator describing the backaction of the measurement on the quantum state for a quantum efficiency $\eta$ and $dw_t$ is a Gaussian distributed variable with variance $dt$ extracted from measurement record normalized appropriately using
\begin{equation}
dw_t=\left(V_t-2\sqrt{\eta  }\mathrm{Tr}[\rho_t \sqrt{\frac{\gamma}{2}} \sigma_Z]\right) dt.
\end{equation}
The probability distribution for the projective outcome is then given by the Born rule $P_{X,Y,Z}(t) = P(y_t|y_0,a,b=X,Y,Z,V_0...V_t)=(\mathrm{Tr}[\rho_t \sigma_{X,Y,Z}]+1)/2$. 
The integrated stochastic master equation provides faithful predictions when  experimental parameters are precisely known from independent calibration under the assumption that the cavity decay rate is much larger than the qubit measurement rate $\kappa\gg \gamma $. 
Fig.2 a. shows two representative trajectories extracted from the measurement records based on the stochastic master equation.


\section{Recurrent Neural Network}
Based solely on a large set of labeled examples $(y_t,y_0,a,b,\{V_\tau\})$ directly extracted from the experimental system, we now demonstrate that the network can be trained to predict the probability $P(y_t|y_0,a,b,V_0...V_t)$ of the observing the measurement outcome $y_t\in\{0,1\}$ given the history of the quantum evolution accessible to the observer, in other words the best knowledge of the qubit wave-function.

We use a Long Short-Term Memory Recurrent Neural Network (LSTM-RNN) \cite{gers2000} schematically depicted in Fig.1b.
These typically consist of a layer of $n$ virtual neurons-like nodes recurrently updated in time. The state of the neuron's layer at a time $t$ is encoded in a $n$-dimensional vector $\vec{h}_t$. It is computed as a weighted linear combination of the neuron's layer state at a previous time $t-1$ combined with the measurement record at a time $t$ and passed through a non-linear activation function  $\phi$  such that $\vec{h}_t=\phi(W_h.\vec{h}_{t-1}+V_t\vec{W_V}+\vec{B}_h)$ where $W$ and $B$ are the weights of the connections between the neurons, and the biases respectively, which are determined during the training stage.
The probability $P_b(y_t)$ of the getting the outcome $y$ given the measurement setting $b$ is computed at each time step as a linear combination of the neuron layer state passed through the activation function given by $P(y_t|y_0,a,b,V_0...V_t)=\sigma(\vec{W_b}.\vec{h}_{t}+\vec{B}_b)$.
The preparation settings $a$ and the initial qubit state (input bit $y_0$) are specified in the initial state of the neuron layer.
The neural network is trained to minimize a loss function $\mathcal{L}$ by strengthening or weakening connections between neuron layers encoded in the weight matrices $W_{h,V,b}$ , as shown in Fig.1b. The cross-entropy loss function $\mathcal{L}^b=-y_T\log{P(y_T|y_0,a,b,V_0...V_T)}-(1-y_T)\log{(1-P(y_T|y_0,a,b,V_0...V_T))}$ is minimized when the prediction $P(y_t|y_0,a,b,V_0...V_t)$ and the distribution of experimental outcomes $y_T$ for a given measurement setting $b$ match.
Crucially, the function implemented by the neural network is differentiable, and therefore the weight matrices can be updated at each iteration of the training  by differentiating the loss-function and applying a gradient-descent minimization step: $W\leftarrow W-\xi \langle\partial \mathcal{L}^b/\partial W \rangle$ where $\xi$ is the learning rate.  The training process ends once the weight matrices $W$ have converged toward a minimum of the loss function. The effectiveness of neural networks lies in their ability to converge toward a minimum of a very high dimensional non-linear loss landscape through gradient back-propagation as illustrated in Fig.1c.


\begin{figure*}
\includegraphics[width=1\textwidth]{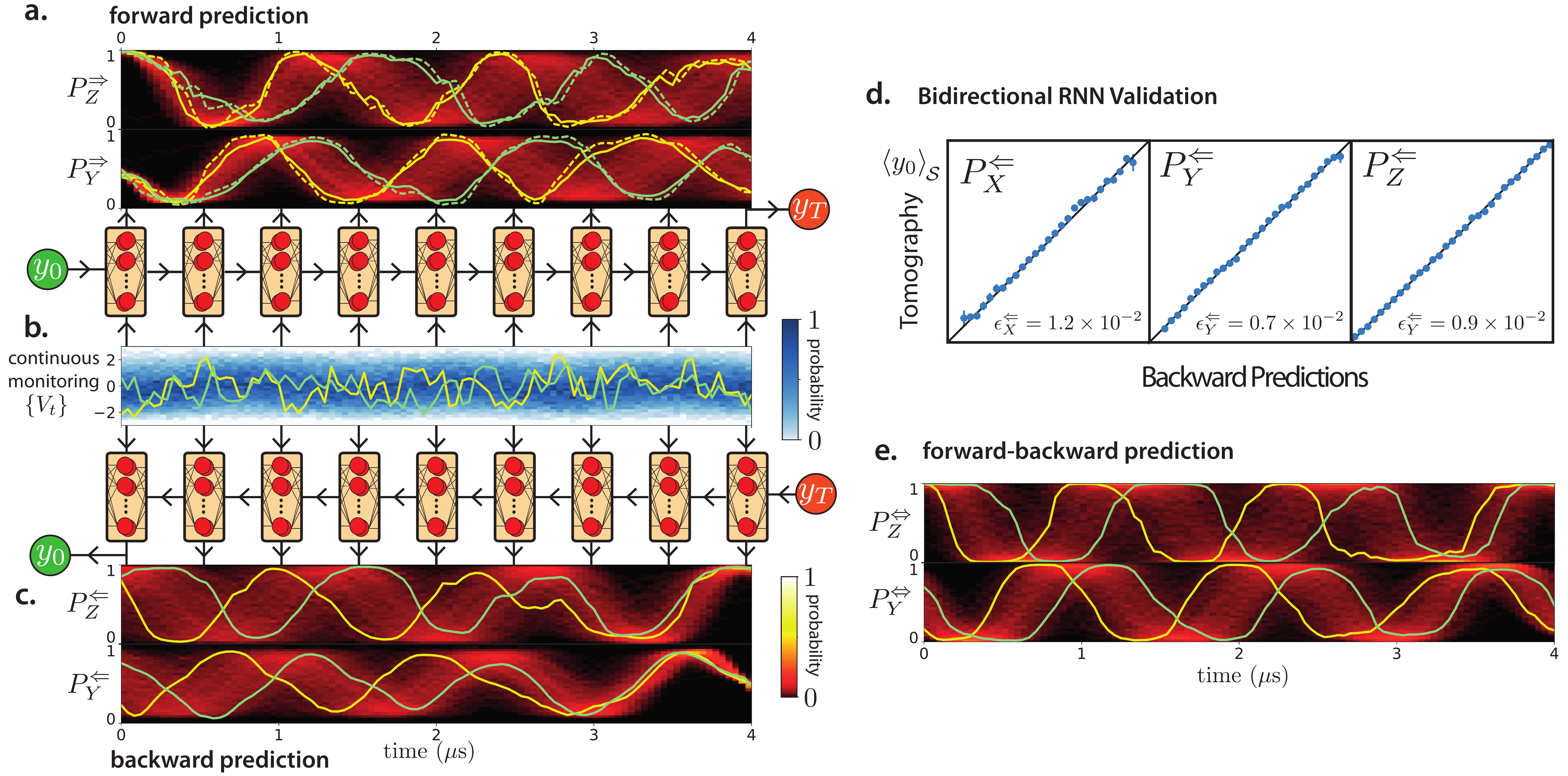}\caption{\textbf{RNN prediction and retrodiction of the quantum evolution}
 \textbf{a}. Red-scale histograms of RNN prediction for the measurement basis $b=Y$ and $Z$ in the driven case beginning from $y_0=1$ in the preparation basis $a=Z$, traces plotted in color show representative instances. \textbf{b}. Blue-scale histogram of the normalized measurement records extracted from the experiment, traces plotted in color show representative instances. \textbf{c}. Red-scale histograms of RNN retrodiction for the same measurement record. {d}. Comparison of the backward RNN prediction with the tomography ---averaged measurement outcome $y_0$. {e}.  Red-scale histograms of smoothed RNN predictions based on the forward-backward analysis given by Eq.(\ref{eq:smoothing}) for the same measurement records.}
 \label{fig:biRNN}
\end{figure*}

\section{Training}

 The Long Short Term Memory recurrent neural network comprises 64 neurons with rectified linear unit activation function. This specific RNN architecture evades the exploding/vanishing gradient problem of standard RNN architecture, improving the learning of long-term dependencies \cite{Hochreiter2001}.  The neural network is implemented with the Tensorflow library \cite{abadi2016} developed by Google and optimized for a Graphics Processing Unit (Nvidia Tesla K80 GPU), which enables a speed up of the training. The data are fed to the network in batches, each containing 1024 measurement records on which a step of the gradient descent is preformed using ADAM optimizer \cite{Kingma2014}. The measurement records is split in two data set. $1.5\times 10^6$ traces are used for the training  and $5 \times 10^5$ randomly chosen traces are used for the evaluation and displayed in the manuscript. The training data can be re-injected several time to the network in order to improve the model accuracy, each of these training cycle corresponds to a training epoch, in practice up to 10 training epochs have been performed.  At each training epoch, the learning rate is lowered from $1\times10^{-3}$ to $1\times10^{-6}$. In order to improve the training robustness, $30\%$ of the neurons are dropped out randomly during the first epoch. The fraction of dropped out neurons is gradually lowered to $0$ with each subsequent training epoch. This method prevents the network from over-fitting and helps the generalization abilities of the model  \cite{Srivastava2014}.
 Note that the training quality does not strongly depend on the details of these parameters.
A key feature of the training is that it can be performed in real-time directly from raw data data collected from the experimental system, the training cycle is $0.8 \mathrm{ms}$ per trace, which is on par with the experimental repetition time. Therefore, the $2\times 10^6$ traces are produced and fed to the RNN in $20\ \mathrm{min}$. 6 preparation settings $(y_0\in\{0,1\}, a\in\{X,Y,Z\})$ and 6 measurement settings $(y_T\in\{0,1\},b\in\{X,Y,Z\})$ are used.  In practice,  we perform the preparation and measurement with the following rotations of the qubit $-$ $R_{\pi/2}^Y$ , $R_{-\pi/2}^Y$, $R_{\pi/2}^X$, $R_{-\pi/2}^X$, $R_0^X$ and $R_{\pi}^X$ $-$ which correspond to the cardinal points of the Bloch sphere. The associated preparation labels $(y_0, a)$ and measurement labels $(y_T, b)$ are then given respectively by  ($y ,X$), ($\bar{y} ,X$), ($y ,Y$), ($\bar{y} ,Y$), ($y ,Z$) and ($\bar{y} ,Z$) with $\bar{y}=1- y$. The total time evolution is varied over 20 values within $4\ \mathrm{\mu s}$, 
$(T\in [0,4])$and  the measurement record $\{V_t\}$ is acquired during the qubit evolution with a sampling time of $40\ \mathrm{ns}$. Once the training achieved, the RNN returns the prediction $P(y_t|y_0,a,b,V_0...V_t)$ which corresponding to the probability of measuring the qubit at a time $t$ along the measurement axis $b=X$, $Y$ and $Z$.

\section{Validation}
Once the RNN is trained, the predictions of the measurement outcomes form an ensemble of trajectories for each of the measurement setting as shown on Fig.\ref{fig:trajectories}b. The prediction of the neural network are in good agreement with the representative trajectories integrated from the stochastic master equation. In this section, we demonstrate that the remaining discrepancies between predictions are in favor of the neural network. The accuracy of the training can be evaluated self-consistently on the evaluation dataset not used during the training. This method has been previously used to benchmark the prediction of the stochastic master equation \cite{Murch2013,Weber2014,Hacohen2016, Ficheux2018}. We select the subset of the trajectories leading to the same prediction $p$ within a small $\delta$ such that $\mathcal{S}_p=\{y_T \text{ such that } P(y_T|y_0,a,b,V_0...V_T)\in[p-\delta,p+\delta]\}$. Fig.\ref{fig:trajectories}c displays the agreement between the ensemble of trajectories ending in $p\pm \delta=0.85\pm0.01$ and the histogram of the final measurement value.  If the prediction is accurate, it should agree with the the final tomographic measurement average on the sub-set $S_p$, defined as $\langle y \rangle_{S_p}=\mathcal{N}_p^{-1}\sum_{y \in \mathcal{S}_p}y$ with $\mathcal{N}_p$ the number of trajectories in $\mathcal{S}_p$, such that $\langle y \rangle_{S_p} =p $. The overall agreement between prediction and the tomography values can be quantified as a relative error $\epsilon=\underset{p}{\sum}  \dfrac{\mathcal{N}_p}{\mathcal{N}}(\langle y \rangle_{\mathcal{S}_p}-p)^2$ where  $\mathcal{N}$ the total number of trajectories. 
As shown in Fig. \ref{fig:trajectories}d, the RNN prediction gives relative error  lower than $10^{-2}$ for all-measurement axis. As a comparison, using the same evaluation data set, the prediction of the stochastic master equation based on the independently calibrated experimental parameters gives a higher relative error along the $Y$ and $Z$ axis. Such a discrepancy can be attributed to small calibration errors and experimental drifts. This self-consistent  evaluation demonstrates the prediction power of the trained RNN and its robustness against calibration errors of physical parameters.

\section{Bidirectional RNN}

RNNs are inherently time oriented; the prediction at a time $t$ $P(y_t|y_0,a,b,V_0...V_t)$ only depends on the measurement record at earlier times.
A common feature used to improve the prediction power of a RNN, for translation application in particular, is to combine the prediction of two RNNs trained respectively forward and backward in time, exploiting the same data in both directions \cite{Schuster1997}. The forward prediction provides the trajectory given the past measurement record $(V_0\rightarrow V_t)$
and the preparation settings $(y_0,a)$: $P^\Rightarrow(y_t)=P(y_t|y_0,a,b,V_0...V_t)$ while the backward prediction provides the trajectory given the "future" measurement record $(V_T\rightarrow V_t)$ played backward and the measurement settings $(y_T,b)$: $P^\Leftarrow(y_t)=P(y_t|y_T,a,b,V_T...V_t)$. 
As shown in Fig.\ref{fig:biRNN}a, the RNN provides an ensemble of backward trajectories. The accuracy of backward prediction are evaluated using the same validation method than the forward prediction, the subset of backward trajectory $\mathcal{S}_{p}$ giving the same prediction $p$ must agree on average with the preparation measurement such that $\langle y_0\rangle_{\mathcal{S}_{p}}=p$. The accuracy of the backward prediction is shown in Fig.\ref{fig:biRNN} b, where the relative error for the preparation settings $X$,$Y$ and $Z$ for the backward predictions are  $\epsilon^{\Leftarrow}_X=1.1\times 10^{-2}$, $\epsilon^{\Leftarrow}_Y=0.9\times 10^{-2}$ and $\epsilon^{\Leftarrow}_Z=0.7\times 10^{-2}$, the overall accuracy is comparable to the forward prediction. Remarkably, the backward and forward predictions do not necessarily agree at a given $t$, indeed these predictions are based on distinct parts of the measurement records. They provide complementary information from the past and future evolution of the system.
Theses predictions can therefore be combined to enhance the knowledge of the quantum state based on the full measurement record. Backward-forward analysis is a well-established postprocessing method with recurrent neural network  \cite{Schuster1997} as well as hidden markov chain methods  \cite{Rabiner1989}. Time-reversal symmetry underlies quantum evolution and exchange the role of state preparation and state measurement  \cite{Aharonov1964}. In a sense, backward-forward analysis naturally translates into quantum regime as the prediction and retrodiction of quantum trajectories  \cite{Gammelmark2013, Campagne2014, Tan2015}. Quantum prediction and retrodiction can be combined based on quantum smoothing techniques  \cite{Tsang2009, Guevara2015} enabling an enhancement of physical parameter estimation  \cite{Rybarczyk2015, Tan2016}. 
The forward and backward predictions can be combined into a smoothed prediction by:
\begin{equation}
P^\Leftrightarrow(y_t)=\dfrac{P^\Leftarrow(y_t)P^\Rightarrow(y_t)}{P^\Leftarrow(y_t)P^\Rightarrow(y_t)+(1-P^\Leftarrow(y_t))(1-P^\Rightarrow(y_t))}.
\label{eq:smoothing}
\end{equation}
As depicted in Fig.\ref{fig:biRNN}c, the smoothed trajectories combine the backward and forward information such that it dismisses the least informative predictions ($P^\Leftarrow(y_t), P^\Rightarrow(y_t)\sim 0.5 $) and strengthen the most informative ones  ($P^\Leftarrow(y_t), P^\Rightarrow(y_t)\sim 0 / 1$). 
By removing ambiguities in the qubit evolution, we access information which is blurred by statistical uncertainties in the standard approach, and we observe an improved temporal resolution on quantum jumps undergone by the qubit. 
The forward-backward analysis demonstrates how bidirectional RNNs naturally combines causal and anti-causal correlations hidden in the measurement records.



\section{Initial State estimation}

The role of the preparation $(y_0,a)$ and measurement $(y_T,b)$ are treated symmetrically in the forward and backward prediction. Hence while the forward RNN predicts the outcome of the final projective measurement, the backward RNN provides an estimation of the initial state of the system given the measurement record. These predictions can be therefore exploited to perform initial state tomography, this task is reminiscent of the enhanced readout discrimination by machine-learning demonstrated in Ref. \cite{magesan2015}. 
For the state estimation, we do not specify the final projective measurement and we initialize the backward network  with a maximally unknown state ($P^{\Leftarrow}(y_T)=0.5$ for $X$, $Y$ and $Z$). Each backward trajectory provides up to 1 bit of information about the initial state \cite{Holevo1973}. Combining this information using maximum-likelihood methods allows for reconstructing the initial state $\vec{P}_0$.  Here, the optimization consists in minimizing the following likelihood function over the probability of the initial state following Ref.  \cite{Six2016}
\begin{equation}
P_0(y_0|a)=\underset{P_0}{\mathrm{argmin}}\left(\sum_n |P_0-
P(y_0|a,V^{(n)}_T...V^{(n)}_0)|^2\right)
\end{equation}
As shown in Fig.4 a, we find an agreement between the initial state estimation and prepation within the $95\%$ confidence interval estimated with bootstraping method. It demonstrates that despite the complicated dynamics, the combination of RNN backward predictions performs as a faithful qubit state tomography. 

\begin{figure}[!ht]
\includegraphics[width=0.5\textwidth]{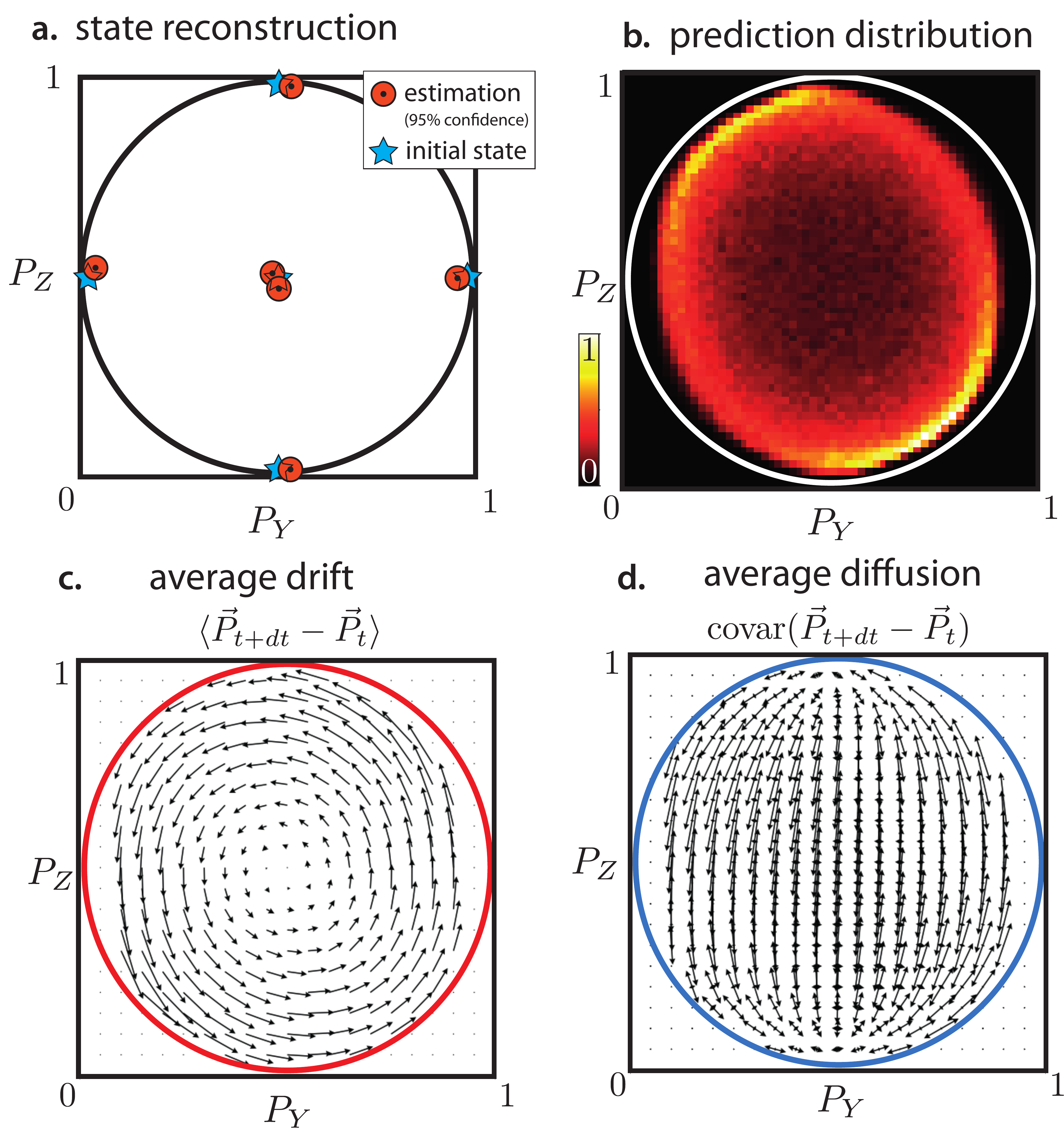}\caption{\textbf{Parameter estimation of the quantum master equation and initial state tomography}  \textbf{a.} - State estimation. Estimation of 6 initial state preparations (red circles) using maximum likelihood estimation on backward RNN predictions ($\sim 20,000$ trajectories each) initialized from an undetermined projective measurement outcome, the circle radius gives the $95\%$ confidence interval extracted from bootstrapping methods. \textbf{b.} Distribution of the RNN predictions in the $Y$ and $Z$ measurement basis for all time. \textbf{c.} Average drift of individual trajectories in the Bloch sphere: The vector map of the averaged evolution of RNN predictions in the $Y$ and $Z$ measurement basis between two consecutive time steps. This map captures the Hamiltonian evolution and the Linbladian dissipation \textbf{d.} Average diffusion of individual trajectories in the Bloch sphere: computed vector map associated with the covariance of the prediction between two consecutive time steps in the $Y$ and $Z$ measurement basis. This map captures the measurement induced backaction.}
\end{figure} 

\section{Parameter estimation}

The trajectories predicted by the trained RNN can be exploited to estimate physical parameters of the experimental system. 
In Fig.4 b, we plot the distribution of the forward RNN prediction in the $Y$,$Z$ plane for all times. This distribution exhibits a tilted ellipse shape within the Bloch sphere (white circle), the great axis of the ellipse is along the $Z$ axis showing that the quantum trajectories tends to collapse toward the poles of the Bloch sphere, corresponding to the pointer states of the measurement operator. In the equatorial plane, the distribution is squeezed, indicating that the quantum state experiences a larger dephasing and  loses purity. 
By performing a statistical analysis of the forward RNN prediction, we are able to reconstruct the physical parameters associated with the stochastic master equation describing the quantum evolution under continuous measurement. The stochastic master equation has two main contributions \cite{Gambetta2008} ; on one hand the dissipative evolution encodes the Hamiltonian evolution along with the decoherence, while on the other hand the measurement back-action describes the update of the quantum state given the stochastic measurement record.
 The dissipative evolution can be extracted from the forward prediction of the RNN by evaluating the average drift of individual trajectories. We compute the ensemble averaged prediction change over intervals of $40\ \mathrm{ns}$,  $d\vec{P}=$ $\langle \vec{P}_{t+1}-\vec{P}_{t} \rangle$ with $\vec{P}_t=(P_X(y_t),P_Y(y_t),P_Z(y_t))$, versus position on the Bloch sphere depicted in Fig. 4c. We observe a drift vector map in the Bloch sphere describing a rotation of the qubit state along the X-axis of the Bloch sphere, corresponding to a Rabi frequency of $\Omega_R/2\pi= 0.82\pm 0.02\  \mathrm{MHz}$. An additional collapse of the state toward the Z-axis corresponds to measurement-induced dephasing rate of $\gamma_\phi= 1.1\pm 0.05\ \mathrm{\mu s^{-1}}$. The measurement-induced disturbance can also be extracted from the prediction of the RNN by evaluating the average diffusion of the individual trajectories \cite{Hacohen2016}. We compute the covariance matrix associated with the prediction change over intervals of $40\ \mathrm{ns}$, $dP^2=\mathrm{covar}(\vec{P}_{t+1}-\vec{P}_t)$. The diffusion vector map is given by the eigenvectors of the covariance matrix weighted by its eigenvalues versus position in the Bloch sphere as depicted in Fig. 4b. This vector map describes the magnitude and the direction of the disturbance induced by the measurement in the Bloch sphere. We observe that the disturbance is maximal along the equatorial plane of the Bloch sphere and vanishes at the poles. From this map, we extract a measurement rate of $\gamma_m= 0.40\pm 0.01\ \mathrm{\mu s^{-1}}$ along the Z-axis of the Bloch sphere. 
The quantum efficiency of our measurement defined as the ratio of the measurement induced dephasing and the measurement rate  gives $\eta= \gamma_m/\gamma_\phi=36\ \%$.
Note that the quantum efficiency is usually challenging to estimate and required several steps of calibrations. The estimated experimental parameters differ sightly from the calibrations which is attributed to residual detuning of the Rabi drive with respect to the qubit frequency. 

\section{Conclusion}
We demonstrate that a recurrent neural network can be trained to provide a model-independent prediction of the outcome of fully general quantum evolution based only on raw observation. The ensemble of predictions can be compared to quantum models such as the stochastic master equation to extract physical parameters without additional calibration. By considering causal and retrocausal evolution, 
we show that initial state tomography can be carried out even for non-trivial quantum evolution.
The black box approach of this work is an illustration of the fact that quantum mechanics is an operational theory, in which states and measurement outcomes can be predicted from raw observation without the mathematical abstraction of a Hilbert space.
The model-agnostic nature of the RNN is therefore readily generalized to larger quantum system. Such networks could excel at finding efficient state representations for larger systems, which could prove useful for real-time modelling, filtering and parameter estimation. The robust, model-independent nature of prediction is a promising tool for the calibration of future quantum processors and will enable characterization of imperfections outside of the scope of the usual approximation, such as correlated errors or non-Markovian noise, and may even be suited for identifying and quantifying effects initially unknown to the experimenter.
\section{Acknowledgements}
We acknowledge M. Devoret, V. Ramasesh, J. Colless and M. Blok for helpful discussions.
LSM acknowledge funding via NSF graduate student fellowships.  This research is  supported in part by the U.S. Army Research Office (ARO) under grant no. W911NF-15-1-0496 and by the AFOSR under grant no. FA9550-12-1-0378.



\bibliography{biblio2}
\bibliographystyle{ieeetr}

\end{document}